\documentclass[twocolumn,amsmath,amssymb,10pt,aps,floatfix,groupaddress]{revtex4-2}
\usepackage{savesym}

\usepackage{graphicx}
\usepackage{dcolumn}
\usepackage{bm}
\usepackage{color}
\usepackage{ulem}
\usepackage{dsfont}
\usepackage[T1]{fontenc}
\usepackage[final]{changes}

\savesymbol{comment} 
\usepackage{comment}

\allowdisplaybreaks
\usepackage[colorlinks,linkcolor=blue,anchorcolor=blue,citecolor=blue,bookmarksnumbered]{hyperref}

\begin{document}

\title{Quantum Multicritical Behavior for Coupled Optical Cavities with Driven Laser Fields}
\author{Yutao Hu$^{1}$}
\author{Yu Zhou$^{2}$}
\author{Wenchen Luo$^{1}$}
\email{luo.wenchen@csu.edu.cn}
\author{Andrea Trombettoni$^{3,5,6}$}
\author{Guoxiang Huang$^{4,7,8}$}
\affiliation{$^{1}$School of Physics and Electronics \& Hunan Key Laboratory of Nanophotonics and Devices, Central South University, Changsha
410083, China}
\affiliation{$^{2}$School of Science and Shenlan College, Jiangsu University of Science and Technology, Jiangsu 212003, China}
\affiliation{$^{3}$Department of Physics, University of Trieste, Strada Costiera 11, I-34151 Trieste, Italy}
\affiliation{$^{4}$State Key Laboratory of Precision Spectroscopy, East China Normal University, Shanghai 200062, China}
\affiliation{$^{5}$SISSA and INFN, Sezione di Trieste, Via Bonomea 265, I-34136 Trieste, Italy}
\affiliation{$^{6}$CNR-IOM DEMOCRITOS Simulation Center and SISSA, Via Bonomea 265, I-34136 Trieste, Italy}
\affiliation{$^{7}$NYU-ECNU Joint Institute of Physics at NYU-Shanghai, Shanghai 200062, China}
\affiliation{$^{8}$Collaborative Innovation Center of Extreme Optics, Shanxi University, Taiyuan, Shanxi 030006, China}

\date{\today}

\begin{abstract}

Quantum phase transitions with multicritical points are fascinating phenomena occurring in interacting quantum many-body systems. However, multicritical points predicted by theory have been rarely verified experimentally; finding multicritical points with specific behaviors and realizing their control remains a challenging topic. Here, we propose a system that a quantized light field interacts with a two-level atomic ensemble coupled by microwave fields in optical cavities, which is described by a generalized Dicke model.
Multicritical points for the superradiant quantum phase transition are shown to occur. We determine the number and position of these critical points and demonstrate that they can be effectively manipulated through the tuning of system parameters. Particularly, we find that the quantum critical points can evolve into a Lifshitz point if the Rabi frequency of the light field is modulated periodically in time. 
Remarkably, the texture of atomic pseudo-spins can be used
to characterize the quantum critical behaviors of the system.
The magnetic orders of the three phases around the Lifshitz point, represented by the atomic pseudo-spins, are similar to those of an axial next-nearest-neighboring Ising model. 
The results reported here are beneficial for unveiling intriguing physics of quantum phase transitions and pave the way towards to find novel quantum multicritical phenomena based on the generalized Dicke model.
\end{abstract}

\maketitle



Critical phenomena is an everlasting subject of interest, both for its
connection with the heart of statistical physics and for its relevance to everyday and technological applications~\cite{Fisher84}. A conceptual point intensely scrutinized focused on how quantum effects modify criticality and to chart quantum phase transitions~\cite{Sachdev}.
In this line of research, a major topic is to characterize and study the quantum counterpart of multicritical points, which is a very active and promising field of research; see, e.g., Ref.~\cite{Fisher10}. In this field, it would be of paramount importance to have a physical setup, highly controllable, in which one can produce and control quantum multicritical points.

A major example of multicritical point is the Lifshitz point (LP) where three phases meet together, i.e., two first-order and a second-order (or two second-order and a first-order) phase transitions intersect in phase diagram. It was shown that the critical behavior around such a point
can be achieved by varying the external parameters, and/or by preparing
mixed compounds or alloys~\cite{LP}.
For instance, a paramagnetic phase, a (anti-) ferromagnetic phase and
a helicoidal phase can meet at a LP, where the second order phase transition
line meshes with the first order one. LPs have been observed in a
variety of condensed matter systems~\cite{LP2,LP3}
and expected to occur in systems of Rydberg atoms \cite{rydberg} and quantum
chromodynamics~\cite{QCD,qcd2}.
Quantum tricritical point (QTP) can be defined as a point
where two lines of first-order and
second-order phase transitions merge, with
properties quite different from those at boundaries of
conventional phase transitions. Remarkable efforts have been made
on QTPs in various systems~\cite{in Fe,in IES, in matels}, and
recently in Dicke models~\cite{xu,Finite,puhan2, huang}.


Dicke model describes typically the interaction between a quantized single-mode light field and an ensemble of two-level atoms~\cite{DCM,Revisit,Mandel1995}, a many-spin version of a quantum Rabi model. In recent years many experimental works have been devoted to systems governed by such a model, including ones by optical cavities~\cite{QED, Optical Cavity,Simulation}, circuit QED~\cite{circuit} and cold atoms~\cite{coldatom}. The Dicke model can exhibit second-order phase transition between superradiant phase (SP) and normal phase (NP)~\cite{Superradiance}. Due to the relative simplicity of theoretical approach, the Dicke model and its extension have become excellent platforms for studying quantum phase transitions, a useful path to simulate and understand the similar phenomena occurring in quantum many-body physics. Although various generalized schemes of Dicke model have been suggested~\cite{Introduction,xu,Finite,Rich,ExploreSB, lead to SB, Dissipation,Hidden in DM,
Hidden in DM 2,Hidden,Controlling,review}, quantum multicritical behaviors predicted theoretically have not been verified by experiment. Hence, finding multicritical points with specific properties and realize their active control in such systems remains a challenging topic up to now.

In this work, we consider a driven system that can be described by an extended
Dicke model, in which a laser field interacts with a gas of identical two-level atoms trapped in $M$ cavities, with each cavity coupled by an arbitrary Zeeman (microwave) field. We show the emergence and manipulation of different quantum critical points (QCPs).
Based on the relation between the order parameter of the radiance
and system parameters, we propose a method
to accurately determine the number and the positions of the QCPs in the phase diagram, which
paves the way to manipulate the QCPs and to detailed study of the criticality in experimental realizations.

Furthermore, we explore the multicriticality of the system when the atom-field interaction is driven periodically by a laser field.  In this case, the system can be described by a correlated model with spin-spin interaction between the cavities. Especially, the QCP can evolve into a LP when the driven is strong. Such a critical behavior is similar to that appearing in an axial next-nearest-neighboring Ising model~\cite{LP3,ANNNI} or quantum Lifshitz model \cite{po}, and the atomic pseudo-spins in each cavity represent the ``magnetic orders'' of different phases around the LP.
Comparing with solid systems, the quantum critical behaviors predicted in the present system, together with its interesting physical properties,
may be easier to observe and can be actively controlled experimentally due to highly tunable of the system.
\added{The results represented here reveal intriguing physics of quantum phase transitions, especially for discovering novel multicritical phenomena in systems linking the Dicke model to quantum many-body systems.}

\section{RESULTS}

{\bf Model}.
We consider a cold atom gas with $N$ two-level atoms, which is trapped equally in $M$ identical high-quality optical cavities and interacts with a quantized single-mode laser field. In each cavity, a microwave field couples the two states of the atoms, with coupling strength $k_j$ for $j$-th cavity ($j=1,2,...,M$). These microwave fields serve as staggered Zeeman fields experienced by the atoms in the cavities, see the Supplementary Material (SM) for details~\cite{appendix}.

The system can be described by an extended Dicke model ($\hbar \equiv 1$) by the Hamiltonian { $H= H_{DM} +H_{EF}$}, with
\begin{eqnarray}
H_{DM} &=&\omega a^{\dagger }a+\sum_{i=1}^{N}\left[ \frac{\delta }{2}\sigma
_{i}^{z}+\frac{G(t)}{2\sqrt{N}}\left( a^{\dagger }+a\right) \sigma _{i}^{x}%
\right] ,  \label{generaldm1} \label{H1}\\
H_{EF} &=&\frac{\omega }{2}\sum_{j=1}^{M}k_{j}\sum_{i=1}^{N/M}\sigma
_{j+M\left( i-1\right) }^{x},  \label{generaldm2}
\end{eqnarray}%
where $H_{DM}$ is the Dicke Hamiltonian; $H_{EF}$ is contributed by the external microwave fields; $a^{(\dagger)}$ is the bosonic operator for the laser field with frequency
$\omega$; $\delta $ is the transition frequency between the two atomic
levels; $\sigma _{i}^{\eta }$ ($\eta =x,y,z; i=1,2,...,N$) are Pauli matrices; and $G$ representing the laser-atom coupling strength is the single-photon Rabi frequency of the laser field, which can be modulated in time by varying the amplitude of the laser field.
The system can be taken to be composed of $M$ subsystems, each containing $N/M$
atoms. We assume, without loss of generality, that the Zeeman coupling strength $k_j$ ($j =1,2,\ldots,M$) is real and can take different signs. This can be
realized by choosing the relative phases of the microwave fields.
For convenience, we take $\omega $ as energy unit, i.e. $\omega =1$.

The multi-cavity (or cavities array) systems, possible to realize in experiments, have been studied to be closely related to strongly correlated systems \cite{multicav}, where each cavity acts as a site in a lattice. We note that too many cavities may be technically challenged, therefore only the systems with a few ($\le 5$) cavities are calculated here. Especially, when the periodic driving is turn on, at most three cavities are considered.

\vspace{2mm}
{\bf Critical points and phase diagram without driving.}
For completeness, we first consider the simple case of a constant coupling strength, $G(t)=g$. Assume that the photon number in the laser field is large, and a mean-field approximation $a \rightarrow\langle a\rangle$ is  applied. In thermodynamic limit ($N\rightarrow \infty$),
the functional of energy-per-atom is given by
$F(\xi) =\xi ^{2}/ \nu -\sum_{j=1}^{M}\sqrt{\delta^{2}+\left( \xi
+k_{j}\right) ^{2}}$, where $\xi =2g \left\langle a \right\rangle / \sqrt{N}$, $\nu =2g^{2}/M$ are the dimensionless order parameter and the laser-atom coupling strength, respectively.
If $\xi =0$ ($\left\vert \xi \right\vert >0 $), the system is in NP (SP).
We use the set, $K=\{k_{1},k_2,\ldots,k_{M}; k_{j}\in \mathbb{R}\}$, 
to represent different Zeeman couplings in different cavities. The ground-state property of the system depends on the choice of $K$.
\begin{figure}[htp]
\centering
\includegraphics[width=8.4cm]{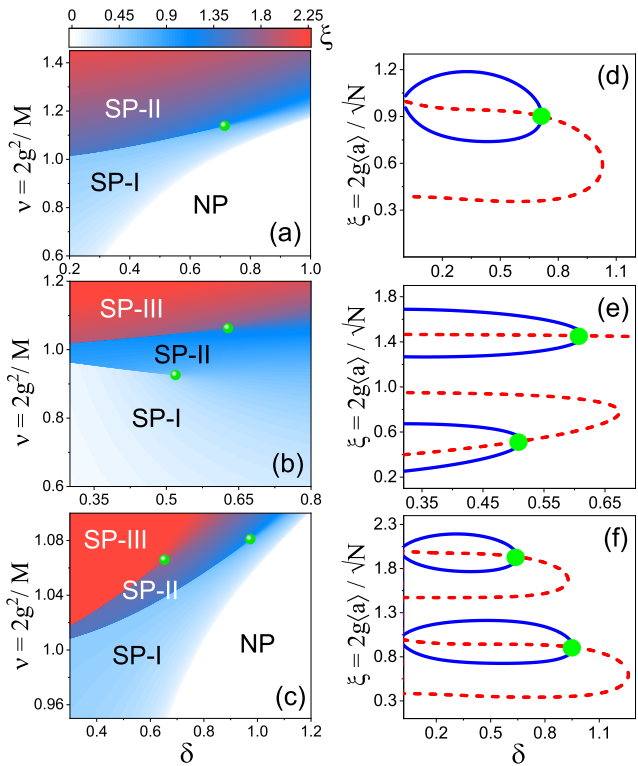}
\caption{
Ground-state phase diagrams as function of $\delta$ (atomic transition frequency) and $\nu=2g^2/M$ (dimensionless laser-atom coupling strength).
(a)~The case for the Zeeman coupling coefficient set $K_3=\{-1,0,1\}$ contains two SPs (i.e. SP-I and SP-II); the green dot in the SP region denotes QCP.
(b)~The same as (a), but with  $ K_4=\{-1.5,-0.5,2,3\}$, where there are three SPs (i.e. SP-I, SP-II, SP-III)  and two QCPs.
(c)~The same as (a), but with  $ K_5=\{-2,-1,0,1,2\}$.  The phase diagram is similar to (a) except that there are three SPs and two QCPs.
The blue solid (red dashed) line in (d), (e), (f) represents the one obtained by solving the equation $\partial \nu/\partial \xi =0$ ($\partial ^{2} \nu/{\ \partial \xi^{2}} =0$). The intersections marked by the green dots in (d), (e), and (f) are the locations of the QCPs for $K_3, K_4, K_5$ [corresponding to the QCPs shown in (a), (b), (c)], respectively.
}
\label{fig1}
\end{figure}

Figs.~\ref{fig1}(a), (b), (c) show the phase diagrams for taking three sets $K_3$, $K_4$ and $K_5$, chosen
respectively as $K_3\equiv \{-1,0,1\}$, $K_{4}\equiv\{-1.5,-0.5,2,3\}$, and $K_5\equiv \{-2,-1,0,1,2\}$, by minimizing the energy-per-atom in the system. We see that, with the increase of $M$ (the number of the cavities), the number of minima of $F(\xi)$ (each minimum corresponds to a SP) is increased, and hence the number of the SP phases is also increased. Moreover,
the phase diagram for the case {of} $K_5$ is similar to that of $K_3$. The former contains three SPs (i.e. SP-I, SP-II, SP-III) and two QCPs, while the later contains two SPs (i.e. SP-I, SP-II) and one QCP.



The positions of the QCPs in the SPs can be obtained by minimizing energy-per-atom, i.e. $\partial F \left( \xi \right) /\partial \xi =0$. From this equation, the parameter $\nu$ can be solved as a function of $\delta$ and $\xi$, $\nu=\nu (\delta ,\xi )\equiv 2\xi  / \sum_{j=1}^{M}(k_{j}+\xi ) [\delta^{2}+(k_{j}+\xi )^{2} ]^{-1/2}$.
Shown in Figs.~\ref{fig1}(d), (e), and (f)  are the results respectively for the Zeeman coupling sets $K_3$, $K_4$, and $K_5$. In the figure, the blue (red dashed) line represents the one obtained $\partial \nu/\partial \xi =0$ ($\partial ^{2} \nu/{\ \partial \xi^{2}} =0$). The intersections marked by the green dots are the locations of the QCPs, corresponding respectively to the QCPs in (a), (b), and (c). For more details, see Sec.~C in SM~\cite{appendix}.
The number of the SP is increased generally when the number of the Zeeman couplings increases. This can be understood by the reason that an addition of new Zeeman couplings can bring a new symmetry breaking in the system.
Furthermore, the number and the location of the QCP can be manipulated by tuning the external fields $K$. Knowing where the QCP is also helpful to determine physical quantities such as the coupling strength $g$ or the energy
gap $\delta $.

\vspace{2mm}
{\bf Multicritiality with a periodic driving.}
We now turn to consider the case that the atom-field coupling is
periodically driven~\cite{Driven}, $G\left(t\right) =g+A\cos \left( \Omega t \right)$. If the driving frequency is so large that we can integrate out all the fast-varied effects, then the effective time-independent Hamiltonian can be obtained for $g,\omega ,\delta \ll \Omega $ (Sec.~D in SM \cite{appendix})
\begin{eqnarray}  \label{h0}
&&h_{0} =\omega a^{\dagger }a+\frac{g\left( a^{\dagger }+a\right) }{\sqrt{N}}%
\sum_{j=1}^{M}J_{j}^{x}+\omega \sum_{j=1}^{M}k_{j}J_{j}^{x}  \label{td-h} \\
&&+\frac{\omega}{2} \frac{A^2}{N \Omega^2 }\left(
\sum_{j=1}^{M}J_{j}^{x}\right) ^{2}+\delta J_{0}\left[ \frac{2A\left(
a^{\dagger }+a\right) }{\Omega \sqrt{N}}\right] \sum_{j=1}^{M}J_{j}^{z},
\notag
\end{eqnarray}%
where $J_{0}$ is the Bessel function of the first kind, $J_j^{\mu}=\frac{1}{2}\sum_i \sigma_{j+M(i-1)}^{\mu}$ are collective spin operators. This Hamiltonian is a combination of the anisotropic Zeeman couplings, the ``spin-orbit'' coupling $(a+a^\dag) J_j^{x,z}$ and the spin-spin interaction $J_i^x \cdot J_j^x$. All the couplings strengths can be
conveniently changed in the highly tunable atomic systems comparing with
the solid-state systems. If there is only one cavity, then no ``spin-spin'' interaction is involved and the Hamiltonian goes back to that in Ref.~\cite{Driven}.

In the thermodynamic limit, the fluctuation can be neglected and the energy of the ground state in the mean-field approach is
obtained by substitutions in  $h_0$: $J_{j}^{z} \to 
\frac{N}{2M} \left( Y_{j}^{2}-1\right)$, $J_{j}^{x} \to 
-\frac{N}{2M}Y_{j}\sqrt{2-Y_{j}^{2}}$, and $a \to  
\sqrt{N }X/\sqrt{2M}$  \cite{appendix,HP}.
This energy functional is highly nonlinear and its minimum can only be found numerically. By minimizing the energy, the order parameters
$Y_j$ and $X=\frac{\sqrt{2M} \xi}{ 2g}$ and the phase of the ground state are obtained.

If there is no staggered Zeeman field, the phase transition is turned to
first-order by the periodic driving at large $\delta$ (or large $g$~\cite{Driven}),
and a QTP appears to join the second-order phase transition remaining at small
$\delta$. If the driving { amplitude $A$ is increased}, the QTP moves in the direction of decreasing $\delta$, until $A$ is so large that all the phase transitions are first-order and the QTP disappears. {In the following, for simplicity but without loss of generality, we focus on the several typical cases where two and three Zeeman fields are included.}

\begin{figure}[htp]
\centering
\includegraphics[scale=0.36]{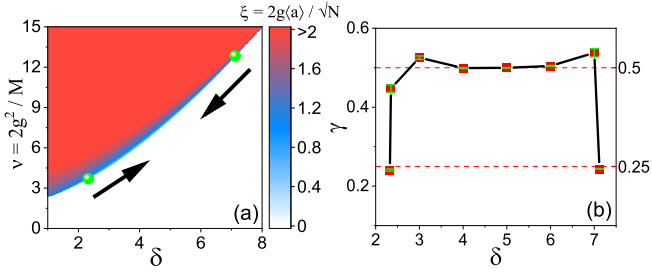}
\caption{(a) The phase diagram of the model with $K=\left\{ -\epsilon, \epsilon
\right\}$ ($\epsilon =1$) and $A/\Omega=0.23$. Two QTPs are marked by the green dots. The QTP at  smaller $\delta$ is induced by the staggered Zeeman couplings while
the other one is induced by the periodic driving $A$. The upper (lower) arrow
represents the direction of the QTP moving when $A$ (or $\epsilon$) is
increased. (b) The critical exponent $\gamma$ and the error bars for the second-order phase transition line between the two QTPs are obtained by power fitting the curve of $\xi$. }
\label{fig2}
\end{figure}

Without the driving, {the case of two staggered Zeeman
fields, i.e. $K=\{-\epsilon, \epsilon \}$,  leads a QTP} joining the first-order phase transition (small $\delta$) to the second-order phase transition (large
$\delta$). Increasing $\epsilon$ moves the QTP in the direction of
increasing $\delta$ \cite{xu}. If both the driving and the Zeeman fields $K_2=\{-1,1\}$ jointly influence the system, the phase diagram [Fig.~\ref{fig2}(a)]
can be understood as the superposition of the
phase diagrams with the driving only { or} with the staggered Zeeman fields only~\cite{appendix}:
At small $\delta $, the first-order phase transition is led by the
superposition of the first-order (from staggered Zeeman fields) and the
second-order (from the driving) phase transitions; When $\delta $ is large,
the opposite occurs; In the intermediate region of $\delta$,
the superposition of two second-order
phase transitions remains the phase transition second-order. As shown in Fig.
\ref{fig2}, the QTP at smaller $\delta$ is due to the staggered Zeeman fields while the QTP at larger $\delta$ is from the driving $A/\Omega =0.23$.
If the staggered Zeeman fields or the driving are increased, the two QTPs move towards each other as shown in Fig.~\ref{fig2}(a), and finally all the phase transitions are first-order and the QTPs disappear once the two QTPs meet.

The critical exponents obtained by fitting the curve of the order parameter $\xi$
\cite{appendix} are $\gamma =0.25$ at the QTPs and $\gamma=0.5$ for the second-order phase transition between the two QTPs.

\begin{figure}[htp]
\centering
\includegraphics[width=8.9cm]{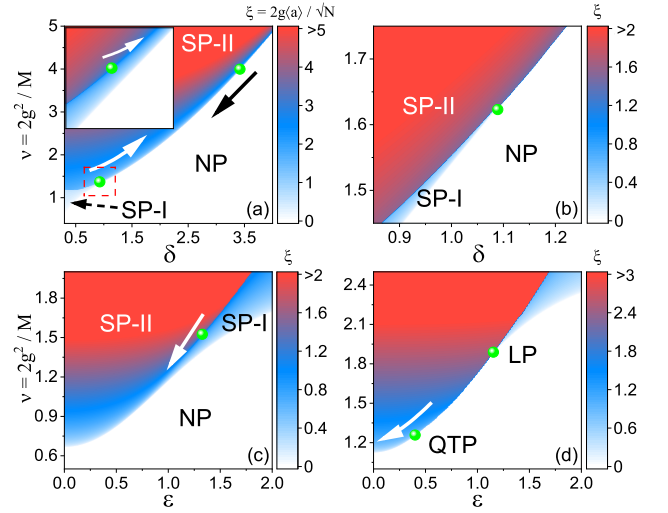}
\caption{The phase diagrams for the model of $K=\left\{ -\epsilon,0,\epsilon\right\}$. (a) A QCP in the
SPs at $\delta=0.85$ and a QTP at $\delta=3.4$ for $A/ \Omega=0.3, \epsilon=1$. The
small region surrounded by the dashed line is zoomed in the inserted panel.
When $A$
increases, the two critical points move in the direction of the arrows. Once the
two points touch, the LP is formed, at about $A/\Omega \approx 0.35$; (b) The
LP is displayed for $A/ \Omega=0.4$. The three phases with
different radiance from low to high are marked as NP, SP-I and SP-II. (c) The
phase diagram with varying $\epsilon$ and fixed $\delta=\omega$,
a QCP separates SP-I and SP-II when the driving is absent. With increase of the driving,
the QCP moves in direction of the arrow. (d) $A/\Omega=0.5$. For $A/\Omega>0.45$, the critical point moves to then moves down
along the NP-SP transition border to evolve to a QTP. Since the SP-I is not significantly
changed, a LP is formed as a triple point, crossed by the first-order SP-I to SP-II and NP to SP-II  transitions, and the second order NP-SP-I phase transition.}
\label{fig3}
\end{figure}

\vspace{2mm}
{\bf Lifshitz point in optical cavities.}
In the case of $K_{3}=\left\{ -1,0,1\right\}$, when the driving is small, the
QCP stays in the SPs to separate SP-I and SP-II, similar to Fig.~\ref{fig1}(a),
while a QTP is generated at large $\delta$ ($\delta=3.3$ for $A/\Omega =0.3$) by the driving, as illustrated in Fig.~\ref{fig3}(a).
The increasing driving pushes the QCP in the SP
region towards the second-order NP-SP transition in the direction of increasing
$\delta$ \cite{appendix} and moves the QTP in the direction of decreasing $\delta$ to form a Lifshitz regime. When $A/ \Omega >0.35$, the QCP and the QTP are pushed
to meet, so that the two first-order phase transition lines connect
together. The SP-I does not disappear at weak coupling $g$, 
a LP is then formed, acting as a triple point that connects SP-I, SP-II and the NP. Two first-order phase transitions (one
between the two SPs, the other between the NP and SP-II) and a second-order
phase transition between the NP and SP-I intersect at the LP, as shown in
Fig.~\ref{fig3}(b). The SP-II has
much stronger radiance than the SP-I, providing the feasibility to observe
the LP  by measuring the {super}radiance.

The {time-dependent} driving mostly affects the SP-II and its related phase transitions,
since the spin-spin interaction is most significant when the spins are polarized.
The phase transition between
spin unpolarized SP-I and NP is unchanged, since $\sum_j J^x_j$ is
small and the spin-spin interaction is not important either.
The reason why the LP appears can be traced from the 
Hamiltonian (\ref{h0}).
Although the atoms do not directly interact with each other, the driven system
is effectively governed by a spin-spin correlated model 
containing physics as rich as strongly correlated systems.
{Its effective Lagrangian is in some sense equivalent to that of the quantum Lifshitz model \cite{po}, if the system is
approximately continuum, due to the existence of $(J^x)^2$.}



\begin{figure}[htp]
\centering
\includegraphics[width=8.2cm]{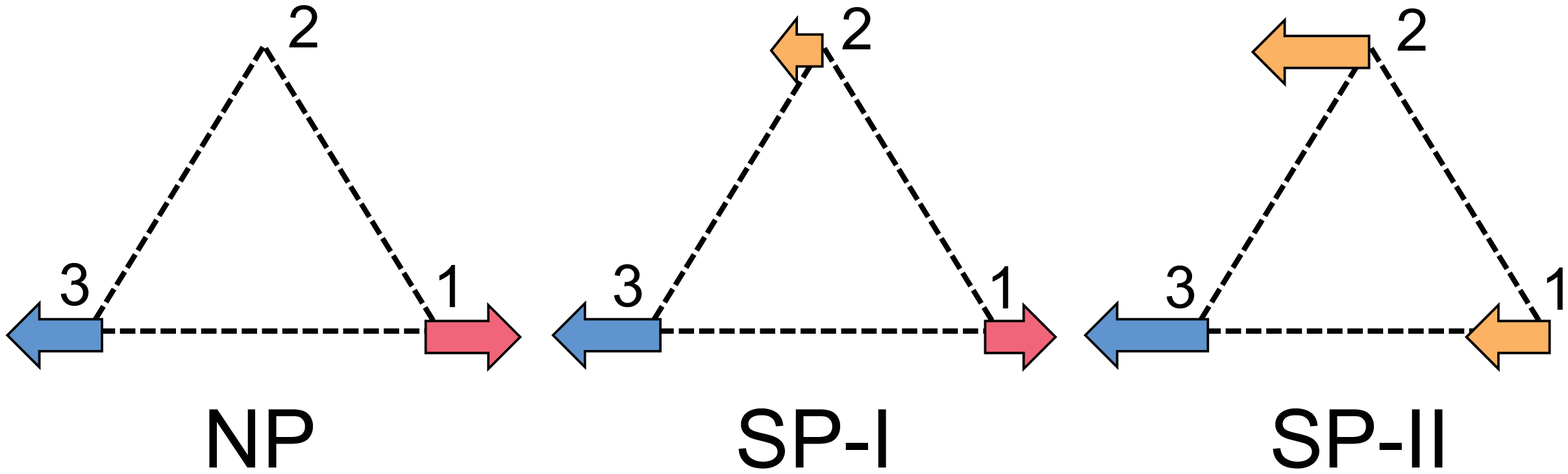}
\caption{In the model with $K_3=\{-1,0,1\}$ and $A/\Omega=0.4$, the
illustration of the $x$ components of the pseudo-spins $\langle J_{1,2,3}^{x}
\rangle$ for the three cavities (1,2,3) of the three phases (NP, SP-I and SP-II) around the LP.
The value of $\langle J_{1,2,3}^{x}\rangle$ are represented by the arrows at
the triangle vertices which represent the three cavities. The NP is an
anti-ferromagnet since $\langle J_{1}^{x} \rangle=-\langle J_{3}^{x} \rangle$ and
$\langle J_{2}^{x} \rangle=0$, the SP-II has a ferromagnetic order:
$\langle J_{1,2,3}^{x} \rangle<0$, and the SP-I has the crossover pseudo-spin textures
with a modulated magnetic order.}
\label{fig4}
\end{figure}

This Hamiltonian is {also} similar to that of the axial next-nearest-neighboring Ising (ANNNI) model
\cite{LP3,ANNNI}, in which the LP exists surrounded by (anti-)ferromagnetic,
paraferromagnetic and modulated phases.
In our case, the two SPs 
and the NP can also be
characterized by the ``magnetic orders'' of the pseudo-spins of the atoms, $\langle J^{x,z}_{1,2,3} \rangle$.
For simplicity, {here} we only monitor $\langle J_{j}^{x} \rangle$ in the three adjacent
phases around the LP, and put $\langle J_{j}^{z}
\rangle$ aside since it is not related to the spin-spin interaction or the
anisotropic Zeeman couplings.
Fig. \ref{fig4} indicates the three different magnetic orders of the three phases: the NP has the
anti-ferromagnetic order, SP-I has the modulated magnetic order, and SP-II
is ferromagnetism. We notice that if the pseudo-spin is more polarized, the
{super}radiance is stronger. The correspondence between the extended Dicke model and
the ANNNI model illustrates why the atomic system can hold a LP surrounded by different magnetic orders.
Besides, for the $K_3=\{-1,0,1\}$ model without driving, the three phases (NP,
SP-I and SP-II) also have these magnetic orders around the QCP. Namely, a
Lifshitz regime including the QCP is perhaps already formed, which however can not
be closed up to a LP due to the lack of the driving induced spin-spin interaction.

We use another variable $\epsilon$ defined in a varied staggered
Zeeman fields $K=\{ -\epsilon, 0, \epsilon\}$  to
plot the phase diagram with a fixed $\delta=\omega$. Without driving, a QCP separating SP-I and SP-II is shown
in Fig.~\ref{fig3}(c). 
The strong staggered Zeeman couplings depolarize the atomic pseudo-spins to weaken the radiance. The SP-I thus appears at large $\epsilon$.
When the driving arises, the QCP is also pushed towards the NP-SP transition
line in the direction of decreasing $\epsilon$ [as the arrow in Fig.
\ref{fig3}(c)]. When $A/\Omega
\approx 0.4$, the point touches the NP-SP transition line to become a LP,
which is intersected by the SP-I to SP-II first-order phase transition and two
second-order phase transitions (NP to SP-I and NP to SP-II). The pseudo-spins
of the three phases in the three cavities, $\langle J_{1,2,3}^{x} \rangle$, are also illustrated by Fig.~\ref{fig4}. If $A$ continues
to increase, the LP remains and a QTP is generated from the LP moving in the
direction of decreasing $\epsilon$ along the NP to SP-II transition line [as the arrow in Fig.~\ref{fig3}(d)]. Meanwhile, the NP to SP-II phase transition
between the LP and QTP turns to be first-order.

\section{Discussion}

In the present study, we have proposed and investigated an extended Dicke model describing a laser-field interacting with an atomic gas trapped in cavities coupled by Zeeman fields. We have shown  how to locate the quantum critical points, which can be manipulated through  tuning the system parameters.
We have demonstrated that, if the atom-field coupling is periodically driven, the system can be mapped to an interacting spin model, which supports rich critical behaviors. Especially, the critical point, close to the normal-superradiant phase transition, forms a Lifshitz regime and finally evolves to a Lifshitz point around which the pseudo-spin of the atoms forms different magnetic orders
due to the competition of the spin-spin interaction and anisotropic Zeeman
couplings. \added{The results reported here are
also helpful for finding novel quantum multicritical phenomena in strongly interacting quantum many-body systems.}

\added{Given that the spin-spin interaction in Eq. (\ref{h0}) is non-local, the system can be easily prevented from the thermalization due to the many-body localization. The energy levels of the long-range interacting system obey the Poisson distribution with disorders\cite{poisson}. Hence, the system may be used to realize the Floquet time crystal \cite{ftc} where the pseudo-spins of the atoms have spontaneous time translational symmetry breaking \cite{unpublished}.}
In addition, several other interesting questions remain open and need to be investigated further. For example, the quantum fluctuation effect of the system and the performance of metrological tasks near the quantum multicritical points exhibited by coupled optical cavities with driven laser fields, etc., are topics deserving to be explored.



{\bf Acknowledgements}\\
{\small
This work was supported by the National Natural Science Foundation of China
under Grant No.~11804396 and Grant No.~11975098. Y. H. acknowledges Yi Jiang
and Muaath Abdulwahab for helpful discussions.
}



\vspace*{0.2cm}

\end{document}